# Single TeraFET Radiation Spectrometer

I. Gorbenko[1], V. Kachorovskii[1,2] and Michael Shur[3,4]
[1]A. F. Ioffe Physico-Technical Institute, 194021 St. Petersburg, Russia
[2]L. D. Landau Institute for Theoretical Physics, Kosygina street 2, 119334 Moscow, Russia
[3] Rensselaer Polytechnic Institute, Troy, NY 12180, USA (e-mail: shurm@ rpi.edu)
[4] Electronics of the Future, Inc., Vienna, VA 22181, USA [3]

*Abstract—* The new TeraFET design with identical source and drain antennas enables a *tunable resonant polarization-sensitive* plasmonic spectrometer operating in the sub-terahertz and terahertz (THz) range of frequencies at room temperature. It could be implemented in different materials systems including silicon. The p-diamond TeraFETs support operation in the 200 to 600 GHz windows.

## I. Introduction

The TeraFETs [1, 2] – the devices using the excitation and rectification of overdamped or resonant plasma oscillations applications implemented in Si [3, 4], GaAs [5, 6], GaN [7], graphene [8], and other materials systems have already been commercialized as tunable and fast detectors of sub-THz and THz radiation. We now show, for the first time that a single symmetric TeraFET or symmetric multiple gate TeraFET structure could be used as spectrometer of THz or sub-THz radiation. The principle of operation uses the phase shift of the radiation induced response at the source and drain contacts of the TeraFET. The phase shift could be controlled either by the incidence angle of the plane wave or by the helicity of the circular polarized radiation. The presented results reveal the new physics of the TeraFET spectrometer and provide the design, characterization and parameter extraction tools for the TeraFET interferometer/spectrometers.

## II. TeraFET Spectrometer Principle of Operation

Fig. 1 showing the TeraFET spectrometer structure illustrates its principle of operation. A THz radiation impinging on the FET coupled to two antennas induces the phase shift $\theta$ between these antennas. This shift depends on the polarization of radiation and geometry of the setup. For circular polarization $\theta$ is nonzero if antennas configuration is asymmetric with respect to direction from the source to drain. For linear polarization the finite phase shift $\theta \propto \sin\varphi$ appears for nonzero incidence angle $\varphi$ (coefficient in this equation depends on details of geometry).

Importantly, this phase shift enters the boundary conditions for the electron fluid in the TeraFET channel

$$U(0) = U_g + U_a \cos(\omega t)$$
$$U(L) = U_g + U_a \cos(\omega t + \theta) \qquad (1)$$

Here $U(0); U(L)$ are the voltages at the source and drain of the channel, respectively, $U_a$ is the THz induced voltage amplitude, $U_g$ is the gate voltage, and $\omega$ is the frequency of the impinging THz radiation. The electron fluid is described by the standard hydrodynamic equations

$$\frac{\partial v}{\partial t} + v \frac{\partial v}{\partial x} + \gamma v = -\frac{e}{m}\frac{\partial U}{\partial x} \qquad (2)$$

$$\frac{\partial v}{\partial t} + v \frac{\partial v}{\partial x} + \gamma v = -\frac{e}{m}\frac{\partial U}{\partial x} \qquad (3)$$

Here $v$ is the velocity of the fluid, $U$ is the gate-to-channel voltage related to the electron concentration in the channel

$$n_s = C(U_g - U)/e \qquad (4)$$

Here $U_g$ is the gate voltage swing (counted form the threshold voltage). The solution of Eqs. (3)-(4) with the boundary conditions given by Eq. (2) yields

$$V = \frac{\beta \omega U_a^2 \sin\theta}{4 U_g |\sin(kL)|^2 \sqrt{\omega^2 + \gamma^2}}, \qquad (5)$$

Here $\beta = 8\sinh\left(\frac{\Gamma L}{s}\right)\sin\left(\frac{\Omega L}{s}\right)$, $k = (\Omega + i\Gamma)/s$,

$$\Omega = \sqrt{\frac{\sqrt{\omega^4 + \omega^2\gamma^2}}{2} + \frac{\omega^2}{2}}; \quad \Gamma = \sqrt{\frac{\sqrt{\omega^4 + \omega^2\gamma^2}}{2} - \frac{\omega^2}{2}}$$

is the wave vector, ω is frequency, Ω is the plasma frequency, $\gamma = 1/\tau$ is the damping rate, $s = \sqrt{eU_g/m}$ is the plasma wave velocity, $e$ is the electron charge, and $m$ is the effective mass.

## III. RESULTS AND DISCUSSION

Figures 2, 3, 4. and 5 show the calculation results for p-diamond FETs, Si NMOS, AlGaN/GaN and InGaAs/InP HEMTs, respectively, for the channel lengths of 25 nm, 65 nm, and 130 nm 9the 250 nm results are also shown for p-diamond). Table 1 lists the parameters used in the calculation.

TABLE I. MATERIALS PARAMETERS

| Material | Parameters | |
|---|---|---|
| | *Effective mass* | *Mobility ($m^2/Vs$)* |
| p-diamond | 0.74 | 0.53 |
| n-Si | 0.19 | 0.10 |
| n-GaN | 0.24 | 0.15 |
| n-InGaAs | 0.041 | 0.8 |

The results show the periodic variation with frequency (adjustable by controlling the phase shift $\theta$ enabling the application as a spectrometer. This spectrometer operates as follows. At each frequency, the gate-to-source voltage will be adjusted till the response is zero at every incidence angle. This yields the values of the frequency satisfying the following condition

$$\sqrt{\frac{\sqrt{\omega_n^4 + \omega_n^2 \gamma^2}}{2} + \frac{\omega_n^2}{2}} = \frac{\pi s}{L} n \qquad (6)$$

Here n=1, 2, 3…. Hence, measuring $\omega_n$ allows for the extraction of the plasma wave velocity and the momentum relaxation time. As seen from Fig. 2 a, the p-diamond FETs should enable room temperature spectroscopy in the sub-THz range. Other materials systems also offer unique capabilities for THz spectroscopy.

ACKNOWLEDGMENT

The work of M. S. S. was supported by the U.S. Army Research Laboratory through the Collaborative Research Alliance for Multi-Scale Modeling of Electronic Materials and by the Office of the Naval Research (Project Monitor Dr. Paul Maki). The work of V. Yu. K. was supported by Russian Science Foundation (grant No. 16-42-01035). The work of I.V.G. was supported by the Foundation for the advancement of theoretical physics "BASIS".

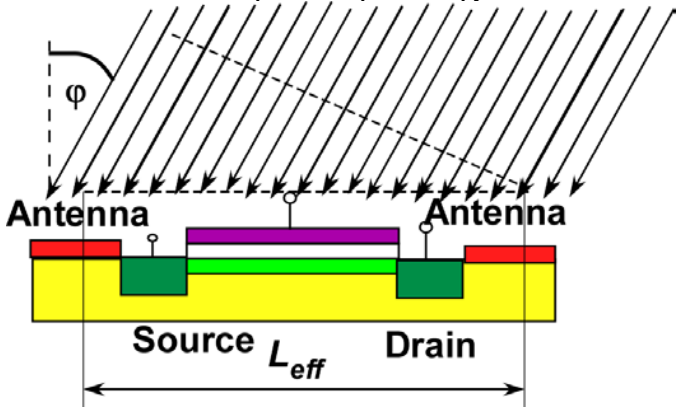

Fig. 1. TeraFET Spectrometer principle of operation.

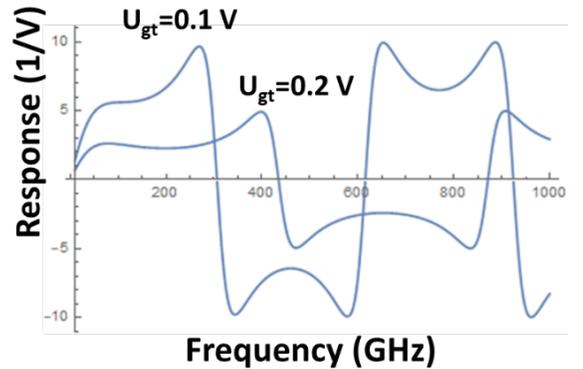

Fig. 2 (a)

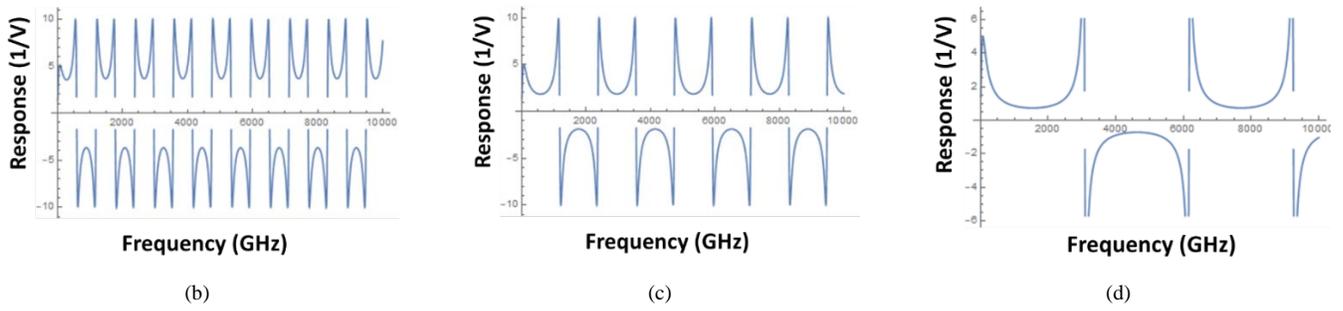

Fig. 2. Diamond TeraFET response normalized to $U_a^2$ for 250 nm (a), 130 nm (b), 65 nm (c) and 25nm (d) channel lengths.

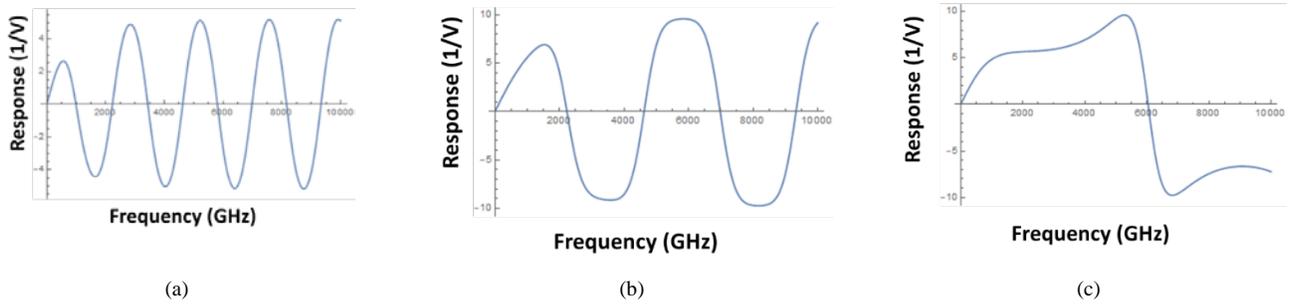

Fig. 3. Silicon TeraFET response normalized to $U_a^2$ for 130 nm (a), 65 nm (b) and 25nm (c) channel lengths.

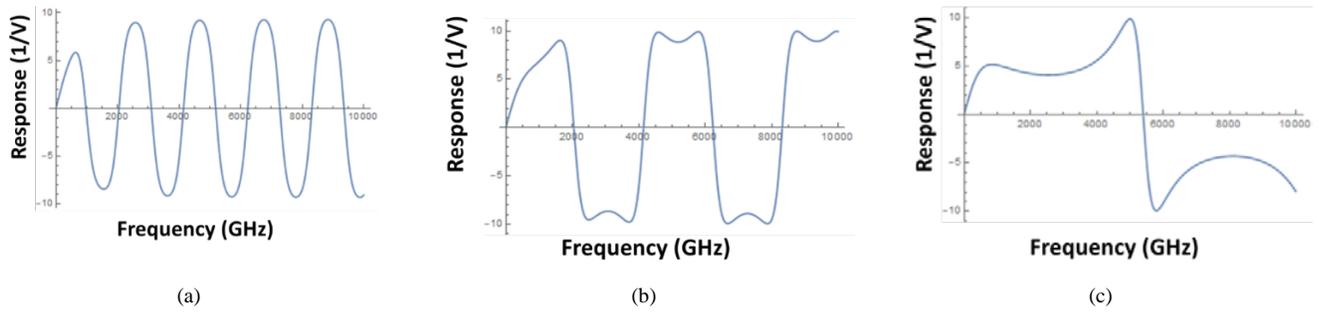

Fig. 4. AlGaN/GaN TeraFET response normalized to $U_a^2$ for 130 nm (a), 65 nm (b) and 25nm (c) channel lengths.

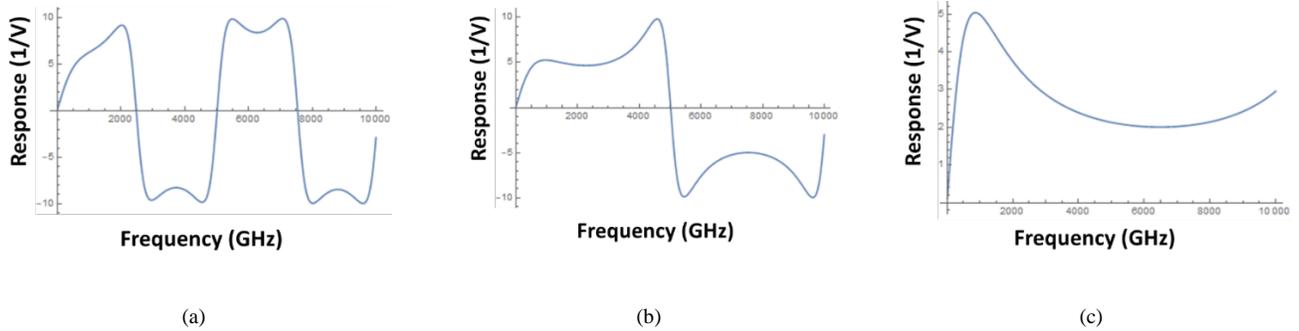

Fig. 5. AlGaAs/GaAs TeraFET response normalized to $U_a^2$ for 130 nm (a), 65 nm (b) and 25nm (c) channel lengths.